\documentclass[twocolumn]{raa}
\usepackage{graphicx,times}
\usepackage{natbib}
\usepackage{amssymb,amsmath}
\bibpunct{(}{)}{;}{a}{}{,}

\usepackage[a4paper=true,driverfallback=dvipdfm,pagebackref=true]{hyperref}

\hypersetup{colorlinks = true, linkcolor = green, anchorcolor = red, citecolor = blue, filecolor = red, pagecolor = red, urlcolor = red}

\mathchardef\mhyphen="2D
\defcitealias{Henriques2015}{H15}

\begin{document}
 \title{Moderate Influence of Halo Spin on Stellar Mass Distributions in Dwarf and Massive Galaxies}

 \volnopage{ {\bf 20XX} Vol.\ {\bf X} No. {\bf XX}, 000--000}
   \setcounter{page}{1}

\author{Yu Rong\inst{1,2}\thanks{corresponding author}, Zichen Hua\inst{1,2}, Huijie Hu\inst{3}
      }

%% Here is an example of three authors come from different institutes.
%% For single author or all the authors from an institute, use "\inst{}" only

   \institute{ Department of Astronomy, University of Science and Technology of China, Hefei, Anhui 230026, China; {\it rongyua@ustc.edu.cn}\\
%% Please give the E-mail address of the author, to whom future correspondence and
%% offprint requests will be sent.
         \and
             School of Astronomy and Space Sciences, University of Science and Technology of China, Hefei 230026, Anhui, China\\
           \and
                  University of Chinese Academy of Sciences, Beijing 100049, China\\
 \vs \no
   {\small Received 20XX Month Day; accepted 20XX Month Day}
}

\abstract{We estimate halo spins for HI-rich galaxies in the Arecibo Legacy Fast Alfa Survey using a semi-analytic approach, examining the relationship between halo spin and stellar surface density. Our findings reveal an inverse correlation in both low- and high-mass galaxy samples, with stellar surface density decreasing as halo spin increases. This trend highlights the pivotal role of halo spin in galaxy evolution and suggests a universal formation scenario: high-spin halos, accompanied by high-spin accreted gas, retain angular momentum, preventing gas from efficiently condensing in the galactic center and thus suppressing star formation. Consequently, weak feedback redistributes gas to the halo outskirts without significant expulsion. The shallower central gravitational potential in high-spin halos promotes outward stellar migration, leading to more extended stellar distributions and lower stellar surface densities.
}

   \authorrunning{Rong et al.}            %author_head in even pages
   \titlerunning{Stellar surface density depends on halo spin}  % title_head in odd pages
   \maketitle
%________________________________________________ sections below
% 
\section{Introduction}           %% first-level sections will be auto-capitalized
\label{sec:1}

In empirical models of galaxy formation, galaxy properties are largely determined by the characteristics of their host dark matter halos. Key processes, including gas accretion, temperature, angular momentum retention, and star formation efficiency, are influenced by these halo properties \citep[e.g.,][]{Rubin10,Rubin15,Lehner14,Fardal01,Keres05,vandeVoort12,Nelson13,Noguchi23,Guo11,Yang12,Behroozi10,Girelli20,Silk77}. For example, halos with shallower potential wells and lower concentrations exhibit more pronounced feedback effects, which leads to reduced star formation efficiencies, lower stellar mass fractions, and more extended stellar distributions \citep{Kravtsov18,Sales22,Hopkins12,Sawala15}. Based on hydro-dynamical simulations, in low-mass halos, gas inflow is dominated by ``cold mode'' accretion, resulting in subsonic flows that create ``hot'' structures with more diffuse stellar distributions \citep{Noguchi18,Noguchi22,Kalita22}. In contrast, gas accretion in massive halos occurs through a ``hot mode'', where supersonic inflows lead to thin, disk-like ``cold'' structures with significant angular momentum \citep{Noguchi18,Hafen22}.

Despite extensive research, the influence of halo spin on galaxy structure and evolution remains incompletely understood. It is generally accepted that the spin of a dark matter halo impacts the stellar distribution in massive late-type galaxies \citep{Mo98,vandenBosch98,Diemand05,Desmond17,Kim13}. However, for low-mass galaxies, high-resolution simulations indicate that halo spin may not play as critical a role as it does in massive counterparts, although it may be essential in forming diffuse stellar distributions in ultra-diffuse galaxies \citep{Yang23,DiCintio19,Rong17a,Amorisco16,Benavides23}.

To date, research on the effects of halo spin on stellar distributions has been inconclusive, particularly in observational studies where halo spin determination poses significant challenges. Observational halo spin measurements have largely focused on small, high-surface-brightness samples \citep[e.g.,][]{Cappellari06,Cappellari13,Wang20,Rong18}, often introducing selection biases. This limitation complicates efforts to accurately investigate the correlation between halo spin and stellar distribution.

Large HI surveys conducted with single-dish telescopes provide a unique opportunity to obtain HI spectra for a large number of galaxies, offering essential dynamical data to estimate halo spin parameters for HI-rich galaxies. Since these samples are selected based on HI column densities rather than stellar characteristics, they present an unbiased approach for examining the relationship between halo spin and stellar distribution.

In this study, we estimate halo spins for HI-bearing galaxies selected from the Arecibo Legacy Fast Alfa Survey (ALFALFA; \citealt{Giovanelli05,Haynes18}) and analyze their relationship with stellar density. Section~\ref{sec:2} describes the data sample and our approach for estimating halo spin. Section~\ref{sec:3} provides a statistical analysis of the correlation between stellar density and halo spin. Our conclusions are summarized and discussed in Section~\ref{sec:4}.

\section{Data}
\label{sec:2}

\subsection{Galaxy sample}

Our sample is drawn from ALFALFA, an extensive HI survey covering roughly 6,600 deg$^2$ at high Galactic latitudes. The final ALFALFA catalog \citep[$\alpha.$100;][]{Haynes18} includes around 31,500 sources with radial velocities below 18,000 km s$^{-1}$, each characterized by properties such as HI spectrum signal-to-noise ratio (SNR), cosmological distance, HI line 50\% peak width of the HI line ($W_{50}$) corrected for instrumental broadening, and HI mass ($M_{\rm{HI}}$). For definitions of these parameters, uncertainties, and estimation techniques, readers are referred to \cite{Haynes18}.

\subsection{Stellar surface density}

ALFALFA sources have been cross-matched with SDSS data \citep{Alam15}. Previous studies by \cite{Durbala20} estimated stellar masses $M_{\star}$ for ALFALFA galaxies with optical counterparts using three methods: UV-optical-infrared SED fitting, SDSS $g-i$ color, and infrared $W_2$ magnitude. We prioritize stellar masses derived from SED fitting, using $g-i$ color estimates only where SED fits are unavailable due to missing UV or infrared data. Discrepancies among these estimates are considered negligible.

To compute stellar densities for the ALFALFA galaxies, we require their effective radii $R_{\rm{e}}$. For approximately 40\% of ALFALFA galaxies, effective radii were measured by \cite{Du19}. For the remainder, we cross-match with the \cite{Simard11} catalog, which provides effective radii for SDSS galaxies. For galaxies without existing radii, we employ the \textsc{SExtractor} software \citep{Bertin96} to estimate $g$-band effective radii following the method of \cite{Du15}.

The stellar density $S_{\star}$ is calculated as $\log S_{\star} = \log M_{\star} + 2.5\log(2\pi R_{\rm{e}}^2)$, with $M_{\star}$ and $R_{\rm{e}}$ given in units of $M_{\odot}$ and kpc, respectively.

\subsection{Halo spin}

To estimate halo spin, we first calculate each galaxy's rotation velocity from its HI spectrum, given by $V_{\rm{rot}}=W_{50}/2/\sin\phi$, where $\phi$ is the HI disk inclination. When direct HI data is unavailable, we estimate $\phi$ using the optical axis ratio $b/a$ from \cite{Durbala20}, calculating inclination as $\sin\phi=\sqrt{(1-(b/a)^2)/(1-q_0^2)}$, setting $\phi=90^{\circ}$ if $b/a \leq q_0$. We use the intrinsic thickness $q_0\sim 0.2$ for massive galaxies, and $q_0\sim 0.4$ for low-mass galaxies with $M_{\star}<10^{9.5}\ M_{\odot}$ (\citealt{Tully09,Giovanelli97,Li21,Rong24}).

To improve rotation velocity accuracy, we exclude galaxies with inclinations $\phi<50^{\circ}$ or low HI SNRs (SNR$<10$) due to higher velocity uncertainties.

For galaxies dominated by velocity dispersion rather than rotation, identified by `single-horned' HI profiles \citep{ElBadry18}, rotation velocity estimates and thus halo spins are unreliable. To classify single- vs. double-horned spectra, we apply the kurtosis parameter $k_4$, following \cite{Hua24}; spectra with $k_4>-1.0$ are considered single-horned. This study focuses on isolated galaxies with double-horned profiles to avoid dispersion-dominated sources.

Halo spin $\lambda_{\rm{h}}$ is then calculated as \citep{Hernandez07}:
\begin{equation}
\lambda_{\rm{h}}\simeq 21.8 \frac{R_{\rm{HI,d}}/{\rm kpc}}{(V_{\rm{rot}}/{\rm km s^{-1}})^{3/2}},
	\label{sam_HI} \end{equation}
where the HI disk scale length, $R_{\rm{HI,d}}$, is derived assuming a thin gas disk in centrifugal balance \citep{Mo98} with an exponential surface density profile:
\begin{equation}
	\Sigma_{\rm{HI}}(R)=\Sigma_{{\rm{HI}},0} {\rm{exp}}(-R/R_{{\rm{HI,d}}}),
\end{equation}
where $\Sigma_{{\rm{HI}},0}$ is the central HI surface density. The total HI mass $M_{\rm{HI}}$ relates to the scale length by:
\begin{equation} M_{\rm{HI}} = 2 \pi \Sigma_{{\rm{HI}},0} R_{{\rm{HI,d}}}^2 \label{HIeq_mass}. \end{equation} 
We introduce the HI radius $r_{\rm{HI}}$, defined as the radius where HI surface density is $1\ \rm M_{\odot}\rm{pc^{-2}}$. $r_{\rm{HI}}$ is estimated using the empirical relation $\log r_{\rm{HI}}=0.51\log M_{\rm{HI}}-3.59$ \citep{Wang16,Gault21}. At $r_{\rm{HI}}$, we set 
\begin{equation} \Sigma_{{\rm{HI}},0} {\rm{exp}}(-r_{\rm{HI}}/R_{{\rm{HI,d}}})=1\ \rm  M_{\odot}\rm{pc^{-2}}. \label{HIeq_3} \end{equation} 
Solving equations~(\ref{HIeq_mass}) and (\ref{HIeq_3}) allows us to compute  $R_{{\rm{HI,d}}}$, enabling robust halo spin estimation for each galaxy in our sample.

%%%%%%%%%%%%%%%%%%%%%%%%%%%%%%%%%%%%%%

\section{Results}
\label{sec:3}

\begin{figure*}
   \centering
   \includegraphics[width=\textwidth, angle=0]{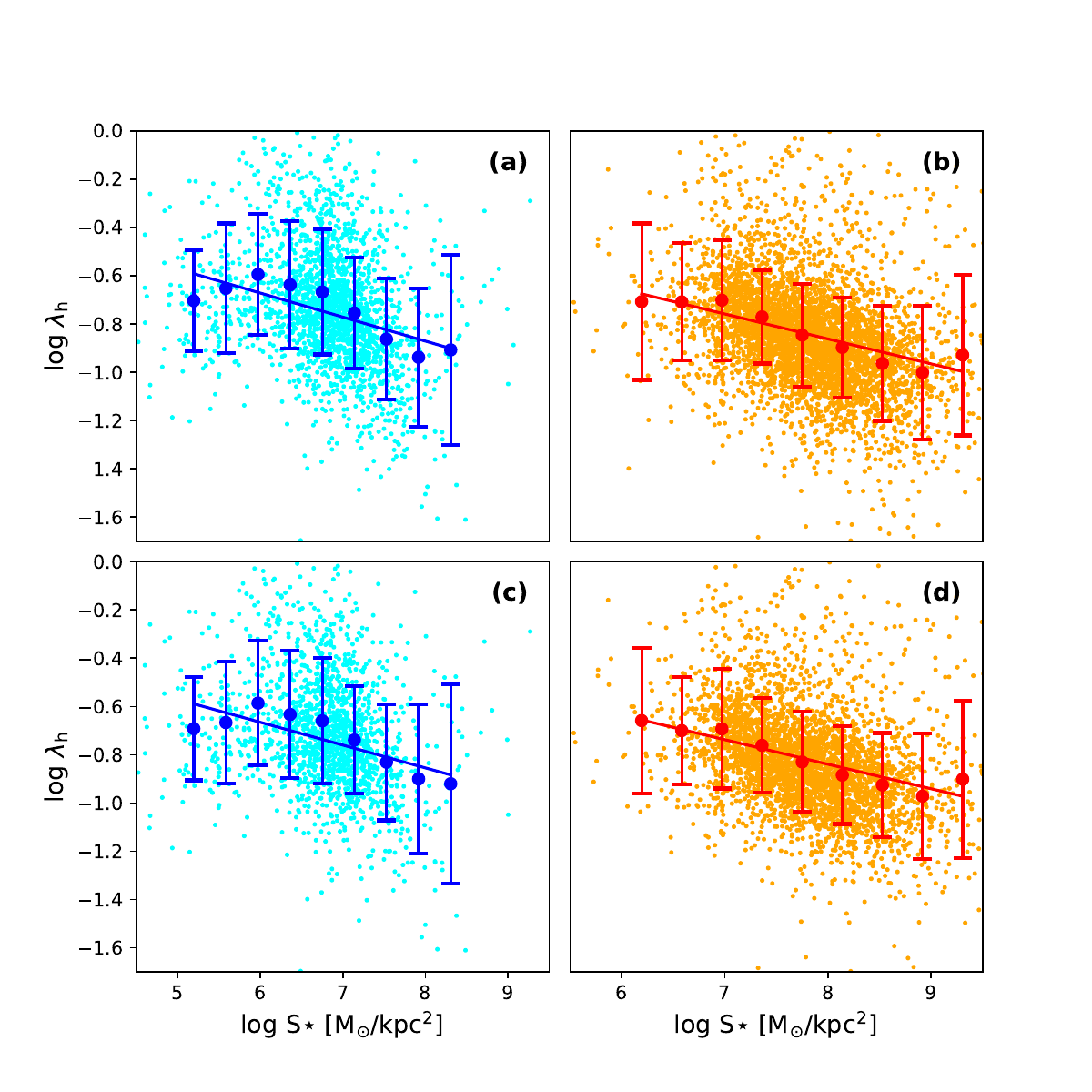}
   \caption{The relationships between stellar surface densities and halo spins for low-mass (left) and massive (right) galaxies. The upper and lower panels correspond to the entire sample and isolated galaxy sample, respectively. Median $\log \lambda_{\rm{h}}$ values with $1\sigma$ error bars are shown in blue and red for bins in $\log S_{\star}$. The best linear fitting results are highlighted by the corresponding lines. }
   \label{fig1}
   \end{figure*}

Our final sample comprises $6,680$ galaxies with stellar masses spanning $10^7$ to $10^{11}$ $\rm M_{\odot}$. Panels~a and b of Fig.~\ref{fig1} illustrate the relationship between $S_{\star}$ and $\lambda_{\rm{h}}$ for the low-mass ($M_{\star}<10^9\ M_{\odot}$) and massive ($M_{\star}>10^9\ M_{\odot}$) galaxies, respectively.The blue and red points with error bars represent the median  $\lambda_{\rm{h}}$ values and their $1\sigma$ uncertainties across different $\log S_{\star}$ bins. A linear fit to these medians yields $\log \lambda_{\rm{h}}= (-0.09\pm 0.09)\log S_{\star}-(0.16\pm 0.59)$ for the low-mass galaxies, and $\log \lambda_{\rm{h}}= (-0.11\pm 0.09)\log S_{\star}+(0.04\pm 0.73)$ for high-mass galaxies. The correlation coefficients of -0.21 and -0.31 suggest a weak correlation in low-mass galaxies and a moderate correlation in high-mass galaxies.

Considering that environmental factors can affect galaxy size and thereby stellar density \citep{Moore96, Mayer01, Mastropietro05, Smith15, Kazantzidis11}, we control for environmental influences by using the galaxy group catalog from \cite{Saulder16}, constructed with a friends-of-friends algorithm based on SDSS DR12 \citep{Alam15} and 2MASS Redshift Survey \citep{Huchra12} data, accounting for biases like the Malmquist bias and ``Fingers of God'' effect. 

We further refine the sample to include only isolated galaxies, defined as those located more than three times the virial radius from any galaxy group or cluster \citep{Rong24a} to minimize effects of tidal stripping and ram-pressure stripping \citep{Mamon04, Gill05, Oman13, Zinger18}. Panels~c and d of Fig.~\ref{fig1} display the relationship between $\lambda_{\rm{h}}$ and $S_{\star}$ for this isolated sample. Linear fits yield $\log \lambda_{\rm{h}}= (-0.08\pm 0.09)\log S_{\star}-(0.20\pm 0.60)$ for isolated low-mass galaxies and $\log \lambda_{\rm{h}}= (-0.11\pm 0.09)\log S_{\star}+(0.03\pm 0.69)$ for isolated high-mass galaxies, with correlation coefficients of -0.18 and -0.30, indicating the persistence of weak to moderate correlations between $S_{\star}$ and $\lambda_{\rm{h}}$ across galaxy masses, independent of environmental effects.

%%%%%%%%%%%%%%%%%%%%%%%%%
\section{Summary and Discussion}
\label{sec:4}

Using a semi-analytic approach, we estimate halo spin parameters ($\lambda_{\rm{h}}$) for HI-rich galaxies in the ALFALFA survey and examine how the stellar surface density ($S_{\star}$) varies with halo spin in isolated low- and high-mass samples. Our analysis reveals a moderate correlation between $\lambda_{\rm{h}}$ and $S_{\star}$ in both galaxy samples, implying that halo spin significantly influences gas cooling, star formation, and feedback processes within galaxies. This finding is consistent with prior results in \cite{Rong24a} but indicates that the correlation between halo spin and stellar surface density is weaker than that between halo spin and HI-to-stellar mass ratios \citep{Liu24}.

As proposed by \cite{Rong24a}, high-spin halos acquire gas with high angular momentum, which resists momentum loss and slows its inward motion and cooling. This process restricts the accumulation of cold gas at the galactic center, reducing the fuel for star formation \citep{Peng20}. Consequently, gas inflow and star formation are more gradual, minimizing supernova-driven feedback and limiting gas expulsion from the halo. Instead, gas is displaced outward within the halo, weakening the central gravitational potential and promoting an extended stellar distribution as stars and dark matter migrate outward \citep{DiCintio17,Yang24}. Our results suggest that this halo-spin-dependent formation scenario may apply universally across galaxy types.

However, the weaker correlation observed between $\lambda_{\rm{h}}$ and $S_{\star}$ compared to that between $\lambda_{\rm{h}}$ and HI-to-stellar mass ratios \citep{Liu24}, suggests that stellar distribution may also be shaped by additional mechanisms beyond halo spin alone, indicating a more complex set of processes at work in galaxy evolution.

Note that we utilize a semi-analytic method to estimate the halo spin, which may introduce relatively large uncertainties. For example, the application of equation~(\ref{sam_HI}) relies on the assumption of spherically symmetric dark matter halo systems, a universally applicable baryonic Tully-Fisher relation with a slope of 3.5, and the same specific angular momenta of cool gas and halo. The assumed geometry and kinematics of the cool gas disk may also be overly idealized. As a result, the estimated spins from equation~(\ref{sam_HI}) are statistically higher than those of dark matter halos in simulations \citep[e.g.,][]{Bett07}. However, since the spins of both galaxy samples with low and high stellar surface densities are likely overestimated, the correlation between $\lambda_{\rm{h}}$ and $S_{\star}$ cannot be attributed to spin overestimation. Nevertheless, this relationship underscores the robust correlation between the two galaxy characteristics, and further observational investigations are warranted to re-examine the moderate correlation between them.

%\normalem

\begin{acknowledgements}

We thank Min He and Wei Du for helps on photometry. Y.R. acknowledges supports from the CAS Pioneer Hundred Talents Program (Category B), and the NSFC grant 12273037, as well as the USTC Research Funds of the Double First-Class Initiative. This work is also supported by the research grants from the China Manned Space Project (the second-stage CSST science projects: ``Investigation of small-scale structures in galaxies and forecasting of observations'' and ``CSST study on specialized galaxies in ultraviolet and multi-band'').

\end{acknowledgements}
  
%\bibliographystyle{raa}
%\bibliography{ms2021-0114}

\end{document}